# An Optimal Lower Bound for Buffer Management in Multi-Queue Switches[*]

Marcin Bienkowski[†]

In the online packet buffering problem (also known as the unweighted FIFO variant of buffer management), we focus on a single network packet switching device with several input ports and one output port. This device forwards unit-size, unit-value packets from input ports to the output port. Buffers attached to input ports may accumulate incoming packets for later transmission; if they cannot accommodate all incoming packets, their excess is lost. A packet buffering algorithm has to choose from which buffers to transmit packets in order to minimize the number of lost packets and thus maximize the throughput.

We present a tight lower bound of $e/(e-1) \approx 1.582$ on the competitive ratio of the throughput maximization, which holds even for fractional or randomized algorithms. This improves the previously best known lower bound of 1.4659 and matches the performance of the algorithm RANDOM SCHEDULE. Our result contradicts the claimed performance of the algorithm RANDOM PERMUTATION; we point out a flaw in its original analysis.

## 1 Introduction

We study *the unweighted FIFO variant of the buffer management problem* introduced by Kesselman et al. [KLM+04]. In this problem, also called *the packet buffering problem* [AS05], we focus on a single packet switching device which has $m$ input ports and one output port. The goal of such device is to forward packets from its input ports to the output port. The burstiness of the incoming traffic motivates the use of buffers that can accumulate incoming packets and store them for later transmission. We assume that all packets are of unit size and that each input port has an attached buffer able to store up to $B$ packets. We consider the unweighted case, in which all packets are equally important.

Time is slotted in the following way. At any (integer) time $t \geq 0$, any number of packets may arrive at the input ports and they are appended to the appropriate buffers. If a buffer cannot accommodate all the packets, the excess is lost. Then, during the time step corresponding to time interval $(t, t + 1)$, the device can transmit a single packet from a single buffer. The key difficulty of this problem is that it is inherently online: the buffer managing algorithm does not know where packets will be injected in the (nearest) future. The goal is to minimize the number of lost packets; the only decision made by the algorithm at time $t$ is choosing the buffer from which a packet is transmitted in step $(t, t + 1)$.

---

[*]Research supported by MNiSW grant number N N206 368839, 2010–2013. A preliminary version of this paper appeared in the proceedings of the 22nd ACM-SIAM Symposium on Discrete Algorithms (SODA 2011). This paper is a restyled version of the manuscript accepted to Algorithmica.

[†]Institute of Computer Science, University of Wrocław, 50-383 Wrocław, Poland. Email: mbi@cs.uni.wroc.pl

In our setting, no information about the future is available to the algorithm, i.e., the decision of an *online* algorithm at time $t$ has to be made solely on the basis of the input sequence up to time $t$. In particular, we make no probabilistic assumptions about the input and assume it is created in a worst-possible manner by an adversary. For analyzing the efficiency of online algorithms, we use competitive analysis [BE98], and — on any input sequence — compare the throughput (the number of transmitted packets) of an algorithm and the optimal *offline* schedule. The supremum (taken over all inputs) of the ratios between these two values is called *the competitive ratio of the algorithm* and is subject to minimization. An algorithm is called $R$-competitive if its competitive ratio is at most $R$. In case of randomized online algorithms, in the definition above, we replace the throughput by its expected value.

The queuing model considered in this paper is typical for input-queued switches or routers, which are the dominant packet switching architecture of the Internet. Moreover, despite the popularity of theoretical research on Quality-of-Service solutions (cf. Section 1.2), most of the current networks (most notably those using IP protocol) provide only "best effort" services, where all packets are treated as equally important and do not have to respect any deadlines. The unweighted variant considered hereby is therefore typical for such networks. Minimizing the packet loss is an important issue, as subsequent packet retransmissions by the standard transport-layer protocols like TCP are quite costly, and may lead to performance degradation. Finally, there is an observable evidence that the network traffic exhibits so-called self-similar properties [LTWW94], which renders analyses based on stochastic queuing theory inapplicable. Hence, it is reasonable to apply the worst-case, competitive analysis rather than stochastic assumptions.

## 1.1 Previous Work

The contributions on this topic fall into three main categories: the results for deterministic algorithms, randomized algorithms and (deterministic) fractional algorithms. The capabilities of the last type of algorithms are extended beyond the standard model: a fractional algorithm may transmit arbitrary fractions of packets as long as the total load transmitted in a single step is at most 1 (see Section 2.1 for a detailed description). It is a straightforward observation (cf. Section 5) that any randomized algorithm can be simulated by a fractional one without increasing its competitive ratio, i.e., the fractional model is "easier" for algorithms and "harder" for the adversary. A relaxed relation in the opposite direction also exists: Azar and Litichevskey showed how to transform a $c$-competitive fractional solution into a deterministic $c \cdot (1 + \lfloor H_m + 1 \rfloor / B)$-competitive one [AL06].

We start by describing results on randomized and fractional models. The currently best lower bound of 1.4659 (holding for arbitrary buffers' sizes $B$ and large $m$) on the competitive ratio of any randomized algorithm is due to Albers and Schmidt [AS05]. In Section 3.2, we highlight the details of their approach and show that in fact it also works for the fractional scenario. Azar and Richter presented the randomized algorithm RANDOM SCHEDULE [AR05], whose competitive ratio is $e/(e-1) + o(1) \approx 1.582$. Azar and Litichevskey [AL06] showed how to encode the fractional variant of the problem as an instance of the online fractional matching in a bipartite graph. By constructing an $e/(e-1)$-competitive solution (based on a natural "water level" approach) for the latter problem, they obtained a fractional $e/(e-1)$-competitive algorithm. In this paper, we refer to it as FRAC-WATERLEVEL. Finally, this bound was falsely claimed to be improved by the algorithm RANDOM PERMUTATION [Sch05] with competitive ratio 1.5 for all values of $B$ and $m$. These competitive ratios were improved for the particular case of two buffers and any $B$, where an algorithm attaining the optimal competitive ratio of $16/13 \approx 1.231$ was given by Bienkowski and Madry [BM08].

As for the deterministic algorithms, the general upper bound holding for all values of $B$ and $m$ was given by Azar and Richter [AR05]. They proved that any *work-conserving* (i.e., transmitting if there is a non-empty queue) algorithm is 2-competitive. They also constructed the lower bound of 1.366 holding for any $B$ and large $m$. Albers and Schmidt [AS05] improved that bound to $e/(e-1)$ for $m \gg B$. They also presented the algorithm SEMI-GREEDY that is 1.889-competitive for large $B$. By taking the algorithm FRAC-WATERLEVEL and applying the fractional-to-deterministic reduction mentioned above, Azar and Litichevskey [AL06] obtained a deterministic $\frac{e}{e-1} \cdot (1 + \lfloor H_m + 1 \rfloor / B)$-competitive algorithm (which we call DET-WATERLEVEL). Again, the results can be improved for particular cases: when $B = 2$, then the SEMI-GREEDY algorithm achieves the optimal competitive ratio of $13/7 \approx 1.857$ [AS05]; when $m = 2$, the optimal ratio $16/13 \approx 1.231$ was achieved by the algorithm SEGMENTAL GREEDY by Kobayashi et al. [KMO08].

The best upper and lower bounds obtained for the general values of $B$ and $m$ are presented in Table 1. Most of the algorithms described above were evaluated experimentally in the paper of Albers and Jacobs [AJ10].

## 1.2 Related Work

The throughput maximization problem considered in the unweighted version in this paper was also studied in a more general context, where packets may have different weights and the goal is to maximize the total transmitted weight. Such an extension tries to capture differences of importance between various data streams. Upon packet injection, the buffer managing algorithm makes a decision whether a packet should be accepted to the buffer or dropped immediately. Afterwards, accepted packets have to be transmitted in FIFO order. Such a problem is non-trivial even for the task of managing a single input buffer [AMRR05, AM03, AMZ03, Zhu04]. A preemptive variant where packets may be dropped from the queue was also studied [And05, AMZ03, EW09, KLM$^+$04, KM03, KMvS05, LP03, MPL04, Zhu04].

Another variant of the buffer management problem is the so-called bounded-delay scenario, where neither the buffer has a fixed capacity nor FIFO order is imposed. Instead, each packet specifies a deadline and it must be either transmitted before the deadline or dropped [AMZ03, BCD$^+$09, BCJ11, CCF$^+$06, CF03, EW07, Haj01, Jeż10, Jeż11, KLM$^+$04, KMvS05, LSS05, LSS07].

For a comprehensive description of these and related models, we refer interested readers to the recent survey by Goldwasser [Gol10].

## 1.3 Our Contribution

The main contribution of this paper (presented in Section 3 and Section 4) is a construction that shows that no fractional algorithm may achieve a competitive ratio lower than $e/(e-1) \approx 1.582$ for any value of $B$ and for large $m$. This result has a few implications (see also Table 1):

- The result is up to lower-order terms optimal; it matches the performance of the fractional algorithm FRAC-WATERLEVEL of [AL06]. It also gives evidence that the reduction of Azar and Litichevskey from the online fractional bipartite matching to the fractional packet buffering was essentially tight. Even though the fractional matching is more general, the competitive ratios achievable for both problems are the same.

|  | Fractional model | Standard model, randomized alg. | Standard model, deterministic alg. |
|---|---|---|---|
| Previous upper b. | $\frac{e}{e-1}$ [AL06] | $\frac{e}{e-1} + o(1)$ [AR05] | $\frac{e}{e-1} \cdot (1 + \lfloor H_m + 1 \rfloor / B)$ [AL06] |
| Previous lower b. | 1.4659 [AS05]* | 1.4659 [AS05] | $\frac{e}{e-1}$ [AS05] (only for $B \ll m$) |
| This paper lower b. | $\frac{e}{e-1}$ | $\frac{e}{e-1}$ | $\frac{e}{e-1}$ |

Table 1: Previously best competitive ratios for the problem for general values of $B$ and $m$. The lower bounds approach values from the table for large $m$. The starred lower bound for the fractional model was not stated in [AS05], but is a straightforward adaptation of the lower bound for the randomized algorithms presented therein (cf. Section 3.2).

- In Section 5, we show a simple reduction showing that any lower bound for the fractional model implies the same lower bound for randomized algorithms in the standard model. Hence, our lower bound improves the currently best lower bound of 1.4659 for randomized algorithms (also holding for any $B$ and large $m$) [AS05]. It is also up to lower-order terms optimal for randomized algorithms as it matches the performance of the algorithm RANDOM SCHEDULE [AR05].

- The lower bound contradicts the claimed competitive ratio of 1.5 of the algorithm RANDOM PERMUTATION [Sch05]. In Section 6, we point out a flaw in the original analysis of this algorithm. The main source of this flaw is neglecting certain types of adversarial strategies; these strategies are actually employed in our lower bound, cf. Section 1.4.

- The lower bound of $e/(e-1)$ for deterministic algorithms given by Albers and Schmidt [AS05] can be applied only when $m \gg B$. Our construction yields the same ratio, but requires only that $m$ is large; $B$ may be arbitrary, e.g., even much larger than $m$. Thus, in contrast to their construction, ours shows that the deterministic algorithm DET-WATERLEVEL [AL06] achieves the asymptotically optimal competitive ratio when both $m$ is large and $B \gg \log m$.

As stated above, our lower bound for randomized algorithms follows simply by the lower bound for fractional algorithms and the relation between randomized and fractional algorithms (cf. Section 5). While for $m \to \infty$ this result cannot be improved, determining the best competitive ratio for smaller values of $m$ remains an open problem. In particular, it is not known whether the fractional and randomized scenarios are equivalent in terms of achievable competitive ratios. In Section 5, we give evidence that it might not be the case. Namely, we show that for $m \geq 3$, no online randomized rounding of a fractional solution is able to preserve the throughput. (A successful randomized rounding for $m = 2$ was given in [BM08].)

## 1.4 Used Techniques

While the formal construction of the lower bound is given in Section 3 and Section 4, here we informally describe its three key ingredients.

First, our adversarial strategy creates a packet injection sequence which at time 0 completely fills all the buffers and can be served losslessly by the optimal algorithm. Moreover, after each injection, all buffers of the optimal algorithm are full. These assumptions are rather standard and we list them only for completeness.

Second, we observe that in some cases delaying packet injections and injecting multiple packets at once incurs a greater packet loss for the algorithm. To give a specific example for a randomized algorithm, assume that the buffer size $B$ is 1 and, at time $t$, the expected number of packets at each buffer is quite low. Then, injecting a packet at times $t$, $t + 1$, and $t + 2$ is more benign to the algorithm than not injecting a packet at time $t$, injecting two packets (to two different buffers, $b_i$ and $b_j$) at time $t + 1$ and injecting a packet at time $t + 2$. At first glance, the former approach tries to incur a loss of the algorithm more aggressively, while the latter gives the algorithm more time to prepare. However, the latter approach can be advantageous to an adversary, because both buffers $b_i$ and $b_j$ become full at time $t + 1$. This can be exploited by injecting a packet at time $t + 2$ to one of these two buffers. On expectation, the incurred loss caused by this injection is then at least $1/2$. Although the observation presented above may seem simple, it was not exploited by previous lower bounds. Moreover, assuming that injecting a single packet at a time constitutes the most cruel adversarial approach is the exact source of the flawed analysis of the algorithm RANDOM PERMUTATION [Sch05].

Third, our construction is an iterative approach that — given an adversarial strategy $S$ for $n$ initially full buffers reducing the average load of the buffers to a specific level — constructs a strategy $S'$ for $n^2$ buffers, which uses the former strategy as a black box. The strategy $S'$ uses only a few more steps (in comparison to the number of buffers) and reduces the average load to even smaller amount than $S$ does. By applying the construction iteratively, in the limit we obtain a strategy for $M$ buffers that reduces the average load of the buffers to $o(1)$ in time $M \cdot (e - 1) + o(M)$. Hence, neglecting the lower-order terms, the throughput of the algorithm on such an injection pattern is $M \cdot (e - 1)$. On the other hand, right after the last injection, the buffers of the optimal solution are still full, and thus its throughput is $M \cdot (e - 1) + M = M \cdot e$. This implies an asymptotic lower bound of $e/(e - 1)$.

## 2 The Model

Throughout the paper, $m$ denotes the number of buffers and $B$ the size of a single buffer. We denote the set of all buffers by $\mathcal{B}$.

At any integer time $t \geq 0$, the adversary may inject an arbitrary number of packets to arbitrary buffers. It may also choose not to inject anything. If a buffer cannot accommodate all the packets, their excess is lost. We assume that packet injection is instant, and therefore we distinguish between the state of buffers at $t^-$ (right before injection at time $t$) and $t^+$ (right after the injection).

Then, during the time step corresponding to time interval $(t, t+1)$, the algorithm may transmit a single packet from the buffer of its choice. We assume that after the last packet injection, the algorithm has sufficient time to transmit all the packets it still has in the buffers.

By a *throughput* of an algorithm $A$ on a sequence $\sigma$, denoted $T_A(\sigma)$, we mean the total number of packets transmitted by $A$. When $A$ is randomized, $T_A(\sigma)$ denotes its expected throughput. For any algorithm ALG, its competitive ratio is defined as $\sup_\sigma \{T_{\text{OPT}}(\sigma)/T_{\text{ALG}}(\sigma)\}$, where the supremum is taken over all possible inputs and OPT is the optimal *offline* algorithm. For randomized algorithms, we assume *oblivious adversaries* [BE98], i.e., the ones that do not have access to the random bits used by the algorithm.

### 2.1 The Fractional Model

It is usually more natural to think about lower bounds for deterministic algorithms. Therefore, in the two subsequent sections, we concentrate solely on a fractional variant of the problem, in which a *deterministic* algorithm may transmit *fractional* amounts of packets. Note that we change solely the capabilities of an online algorithm, which means that: (i) the injected packets are still integral and can be injected at integer times only, (ii) an optimal offline solution to which our solution is compared is chosen among integral solutions. We call the amount of packets in a buffer the *load* of this buffer. For a subset of buffers, the load of these buffers is the sum of the respective buffers' loads, and the total load is the load of all buffers. Furthermore, in each step the load transmitted from the buffers is at most 1. We show relations between fractional and randomized algorithms in Section 5.

By Lemma 5.1 of [AL06], without loss of generality, we may assume that a fractional algorithm is *work-conserving*, i.e., in each step it transmits a load of 1 if it has it, and its total load otherwise. We silently make this assumption in our proofs.

## 3 Basic Building Blocks

In our construction, for simplicity of the description, we assume that $B = 1$. However, the presented approach is valid with virtually no changes for other values of $B$: we only have to replace an injection of a single packet by injection of $B$ packets to the same buffer and replace each step by $B$ consecutive steps.

An elementary building block of the constructed input sequence is *an action*. It is parametrized by a subset of buffers $A$ and a positive integer $a$ called length. When we say that the adversary executes action $(A, a)$ at time $t$, we mean that: (i) nothing is injected during time $(t, t + a)$ and (ii) at time $t + a$, the adversary chooses $a$ buffers with the maximal load among buffers from $A$ and injects a packet to each of these buffers. Ties are broken arbitrarily. We call set $A$ *active* in time $(t, t+a]$ or active for this action, and we say that such action *operates* on set $A$. The considered action *starts* at time $t$ and *ends* at time $t + a$.

### 3.1 Canonical Strategies

A brief characterization of the inputs we construct is as follows. First, at time 0, the adversary injects a packet to each of $m$ buffers; this is called the *initial injection*. Second, at time 0, it *executes* a *canonical (adversarial) strategy*. Such a strategy is a sequence of actions, where for each consecutive pair of actions, the latter starts exactly when the former ends. (For technical reasons, which become clear when we describe our recursive construction, the initial injection is not the part of the canonical strategy.) Furthermore, we define the *length* of a canonical strategy as the sum of its actions' lengths; note that the length of a strategy starting at time 0 coincides with the last time at which packets are injected. We say that something happens *during* the strategy of length $\ell$ executed at $\tau$ if it happens within time interval $(\tau, \tau + \ell]$.

In our construction, we consider canonical strategies only. Canonical strategies are neat to analyze, because the optimal algorithm can serve the corresponding inputs losslessly.

**Lemma 1.** *For a canonical adversarial strategy of length $\ell$ preceded by the initial injection, the throughput of the optimal offline algorithm is $m + \ell$.*

*Proof.* Let $(A_1, a_1), (A_2, a_2), \ldots, (A_k, a_k)$ be the canonical strategy. For any $1 \leq j \leq k$, during $a_j$ time steps of action $j$, OPT transmits packets from these $a_j$ buffers to which a packet is injected at time $\sum_{i=1}^{j} a_i$ (i.e., at the end of action $j$). In effect, during the first $\sum_{i=1}^{k} a_i = \ell$ steps, OPT transmits a packet in each step and does not lose any. Finally, at time $\ell^+$, the buffers of OPT are still full, and hence it may transmit all $m$ packets in time $(\ell, \ell + m)$. □

By the lemma above, we may use a canonical strategy to show a lower bound on any algorithm in the following way. We show that after a given strategy $S$ of length $\ell$ is executed against any (work-conserving) algorithm ALG, the total load of ALG at time $\ell^+$ is at most $c$. Then, during the first $\ell$ steps, the load transmitted by ALG is at most $\ell$, and afterwards it is at most $c$. On the other hand, by Lemma 1, the throughput of OPT is $\ell + m$. Hence, strategy $S$ implies that the competitive ratio of any algorithm is at least $(\ell + m)/(\ell + c)$. Therefore, our goal is to construct an adversarial canonical strategy that reduces the total load of any algorithm as quickly as possible.

In our construction, for a given number of buffers $m$, we give a strategy where the lengths of the corresponding actions are fixed (and known to the algorithm). However, the choice of active sets is algorithm dependent.

### 3.2 Uniform Strategies

The most straightforward canonical strategies are of the form $(A, 1), (A, 1), \ldots, (A, 1)$, where $A$ is a fixed set of buffers. In other words, when executing such a strategy, the adversary injects a packet at the end of each step into the most populated buffer from set $A$. We call such strategies *A-uniform*.

We use $(A, a) \times k$ as a shorthand for the strategy consisting of $k$ consecutive actions $(A, a)$. The behavior of algorithms on $\mathcal{B}$-uniform strategies (recall that $\mathcal{B}$ is the set of all buffers) was investigated previously by Albers and Schmidt [AS05]. Hereby, we show a slightly more general result, which becomes useful later: we consider $A$-uniform strategies for arbitrary sets $A$. Note that during an execution of such a strategy, the algorithm may transmit load also from buffers outside of $A$.

**Definition 2.** *For any subset $A$ consisting of $n$ buffers and a real $\beta > 0$, the strategy $S_0(\beta, A)$ is the canonical uniform strategy $(A, 1) \times \lceil \beta n \rceil$.*

**Lemma 3.** *Fix a real $\beta > 0$, any subset of $n$ buffers $A$, and let $\ell$ be the length of the strategy $S_0(\beta, A)$. Assume that at a step $\tau$, all buffers of $A$ are full and the adversary starts to execute the strategy $S_0(\beta, A)$. Let $H$ be the load transmitted during $S_0(\beta, A)$ from buffers not in $A$. Then the load in $A$ at time $(\tau + \ell)^+$ is at most $H + n \cdot e^{-\beta} + 1$.*

*Proof.* For simplifying the notation, we assume that $\tau = 0$. For any integer $t \geq 0$, let $C_t$ be the load of $A$ at time $t^+$, and for $t \geq 1$, let $a_t \in [0, 1]$ be the load transmitted from $A$ in step $(t - 1, t)$.

As a packet is injected at each time, the algorithm is able to transmit load 1 in each step, i.e., $H = \sum_{t=1}^{\ell}(1 - a_t)$.

We establish a recursive relation for $C_t$. Clearly, $C_0 = n$. Fix any $t \in \{1, \ldots, \ell\}$. At time $t^-$, the load in the most populated buffer of $A$ is at least $(C_{t-1} - a_t)/n$, and thus the load in the remaining buffers is at most $(C_{t-1} - a_t) \cdot (1 - 1/n)$. Due to the injection, the load in $A$ becomes $C_t \leq (C_{t-1} - a_t) \cdot (1 - 1/n) + 1 \leq (C_{t-1} - 1) \cdot (1 - 1/n) + (1 - a_t) + 1$. Hence, $C_\ell \leq (n - 1) \cdot (1 - 1/n)^\ell + \sum_{t=1}^{\ell}(1 - a_t) + 1 < n \cdot e^{-\ell/n} + H + 1$. As $\ell \geq \beta n$, the lemma follows. □

We remark that the previously best lower bound, due to Albers and Schmidt [AS05], was to use the strategy $S_0(\beta, \mathcal{B})$ with a numerically optimized $\beta$ and large $m$. The best choice is $\beta \approx 1.1462$, which for $m \to \infty$ causes any algorithm to have competitive ratio at least 1.4659.

### 3.3 Delayed Injections

As mentioned in Section 1.4, the crucial observation behind our construction is that sometimes it is beneficial to delay injections first and then to inject more than one packet at once. To give a more specific example consider the following canonical strategy for $m = n^2$ buffers:

$$P = (\mathcal{B}, n), (A_1, 1) \times 1, \ (\mathcal{B}, n), (A_2, 1) \times 2, \ \ldots, \ (\mathcal{B}, n), (A_n, 1) \times n \ ,$$

where $A_j$ is the set of $n$ buffers to which packets were injected during the action $(\mathcal{B}, n)$ immediately preceding actions $(A_j, 1) \times j$. This example serves as a basis for our construction; below we shortly and informally explain its design rationale.

Consider a $\mathcal{B}$-uniform strategy. A closer examination (see the proof of Lemma 3) reveals that it decreases the total load by a multiplicative factor in each round. Hence, it is quite effective when buffers are full, but becomes less efficient when the average load drops. To alleviate this problem, consider replacing $n$ steps of the uniform strategy (i.e., the sequence $(\mathcal{B}, 1) \times n$) by a single $n$-step action $(\mathcal{B}, n)$. Although such replacement only further worsens the effectiveness of the adversary, the advantage of having $n$ full buffers (denoted $A_j$) after the execution of $(\mathcal{B}, n)$ outweighs this loss. Concretely, the adversary may execute the uniform strategy $(A_j, 1) \times j$ on the buffers of $A_j$. As these buffers are initially full, the uniform strategy reduces the load in $A_j$ quite rapidly. The adversarial phase $(\mathcal{B}, n), (A_j, 1) \times j$ is repeated for increasing values of $j$. This balances the influence of $(\mathcal{B}, n)$ and $(A_j, 1) \times j$, assuring that the average load inside $A_j$ after a single phase is roughly the same as the load in $\mathcal{B} \setminus A_j$.

The analysis given in the next section shows that already this simple strategy $P$ incurs a lower bound of 1.5589 on the competitive ratio of any fractional algorithm. How can it be further improved? Let us compare $P$ with the $\mathcal{B}$-uniform strategy of the same length. By Lemma 1, the throughput of OPT on both strategies is the same, and therefore the key factor determining the efficiency is the total load of an online algorithm when these strategies end. It can be computed that its total load at that time is at most $0.104\,m$ for the strategy $P$ and at most $0.223\,m$ for the $\mathcal{B}$-uniform strategy.

Now, we take a closer look at a single part $(A_j, 1) \times j$ of the strategy $P$. It is focused entirely on decreasing the load in $A_j$. (The algorithm might transmit the load from other buffers as well, but as we implicitly show later, this is not beneficial.) As the strategy $P$ is more efficient in decreasing the load than the uniform strategy, we could replace the sequence $(A_j, 1) \times j$ by a $P$-type strategy of length roughly $j$ operating on the set $A_j$. Furthermore, we could perform such a replacement recursively! Such top-down replacements are viable, but troublesome to define. Therefore, in our construction, we apply a bottom-up approach: we fix $n$, start from a strategy for $n$ buffers, and using it as a black box, we build a better strategy for $n^2$ buffers. By applying such a transformation iteratively, in the limit we obtain a desired construction, incurring a lower bound of $e/(e-1)$.

## 4 Lower Bound Construction

Let $n$ be a parameter of our construction, which will be specified later. Let $n_0 = n$ and $n_{q+1} = n_q^2$ for any integer $q \geq 0$. In Definition 2, we have already specified strategies $S_0(\beta, A)$ operating on any set $A$ of $n_0$ buffers. Now, we iteratively generalize this definition to any set of $n_q$ buffers.

**Definition 4.** *Fix an integer $q \geq 0$. For any subset $A$ of $n_{q+1}$ buffers, and any $\beta \in (0, 1]$, let*

$$\begin{aligned}
S_{q+1}(\beta, A) = \ &(A, n_q), \ S_q(1/n_q, A_1), \\
&(A, n_q), \ S_q(2/n_q, A_2), \\
&\quad \vdots \\
&(A, n_q), \ S_q(\lceil \beta n_q \rceil / n_q, A_{\lceil \beta n_q \rceil}) \ ,
\end{aligned}$$

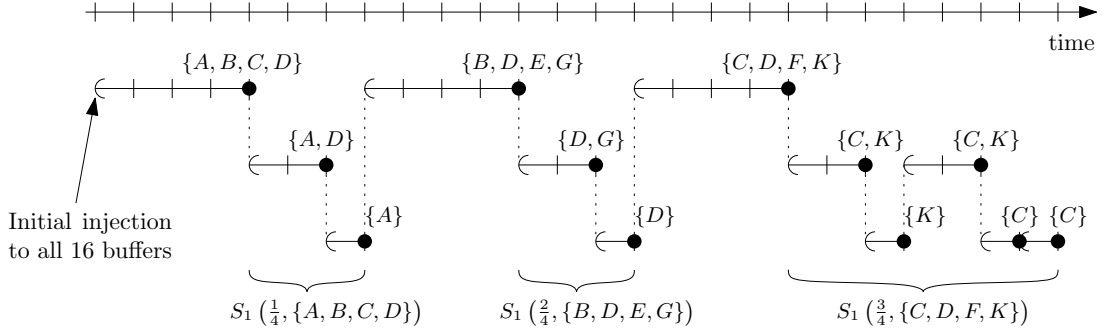

Figure 1: An example strategy $S_2(3/4, \{A, B, \ldots, P\})$ operating on 16 buffers identified with capital letters. Sets on the picture denote buffers to which the adversary injects packets at dot-marked times. The strategy $S_2(3/4, \{A, B, \ldots, P\})$ uses three substrategies: $S_1(1/4, \{A, B, C, D\})$, $S_1(2/4, \{B, D, E, G\})$ and $S_1(3/4, \{C, D, F, K\})$.

where $A_j$ is the set of $n_q$ buffers to which packets were injected during the action $(A, n_q)$ immediately preceding $S_q(j/n_q, A_j)$.

We note that when $m = n^2$, the definitions of the strategy $S_1(1, \mathcal{B})$ and the strategy $P$ from Section 3.3 coincide.

We say that the strategy $S_q(\beta, A)$ is a *routine $S_q(\beta)$ operating on set $A$*. This decouples the actual structure of the adversarial strategy from the sets used. In particular, it is helpful to interpret $S_{q+1}(\beta)$ as a (program) routine, which calls subroutines $S_q(1/n_q), S_q(2/n_q), \ldots, S_q(\lceil \beta n_q \rceil / n_q)$ (cf. Definition 4). We emphasize that the called subroutines do not depend on the actual choices of the algorithm. (On the other hand, *the sets* these routines are operating on depend heavily on the algorithm.) In particular, the length of the routine $S_q(\beta)$ does not depend on the algorithm.

By Definition 4, if a routine operating on set $A$ calls a subroutine operating on set $A'$, then $A' \subseteq A$ and all buffers from $A'$ are full at the beginning of the subroutine. Hence, during strategy $S_q(\beta, A)$ the adversary focuses entirely on the buffers from $A$, ignoring the remaining ones. (Note that the algorithm is not bounded in this way: it may transmit the load from an arbitrary buffer.) An example strategy is given in Figure 1.

The ultimate goal of the remaining part of this section is to analyze the strategy $S_q(1, \mathcal{B})$ on $m = n_q$ buffers, showing that — for appropriately chosen values of $n$ and $q$ — it takes time $(e - 1) \cdot m + o(m)$ and reduces the total load of any online algorithm to $o(m)$. This would immediately imply the lower bound of $e/(e-1)$.

### 4.1 Bounding the Length of $S_q(\beta)$

We start with bounding the length of our recursively defined strategies. As noted above, these lengths depend neither on the algorithm nor on the used sets. For any integer $q \geq 0$ and real $\beta \in (0, 1]$, we denote the length of the adversarial routine $S_q(\beta)$ by $T_q(\beta)$.

We start with the following technical lemma that is used extensively later.

**Lemma 5.** *For any non-negative integers $n > 0$, $q$, $k$ and any positive real $\beta$, the following inequality holds:*

$$\sum_{j=1}^{\lceil \beta n_q \rceil} \left( \frac{j}{n_q} + \frac{2q}{n} \right)^k \leq \frac{n_q}{k+1} \cdot \left( \beta + \frac{2(q+1)}{n} \right)^{k+1} .$$

*Proof.* We bound the sum by an integral, obtaining

$$\sum_{j=1}^{\lceil \beta n_q \rceil} \left( \frac{j}{n_q} + \frac{2q}{n} \right)^k \leq \int_1^{\beta n_q + 2} \left( \frac{j}{n_q} + \frac{2q}{n} \right)^k \, dj \leq n_q \cdot \frac{((\beta n_q + 2)/n_q + 2q/n)^{k+1}}{k+1}$$

$$\leq \frac{n_q}{k+1} \cdot \left( \beta + \frac{2(q+1)}{n} \right)^{k+1} ,$$

where the last inequality follows by $n_q \geq n$. □

**Lemma 6.** *Fix integers $n > 0$, $q \geq 0$, and a real $\beta \in (0, 1]$. Then,*

$$T_q(\beta) \leq n_q \cdot \left( \frac{1}{n} + \sum_{k=1}^{q+1} \frac{(\beta + 2q/n)^k}{k!} \right) .$$

*Proof.* The proof follows by induction on $q$. For $q = 0$, the right hand side of the inequality is $\beta n + 1$, and thus the induction basis holds by Definition 2.

We assume that the bound holds for $q$, and prove it for $q + 1$. Using Definition 4 first and then the inductive assumption, we obtain

$$T_{q+1}(\beta) = \sum_{j=1}^{\lceil \beta n_q \rceil} \left( n_q + T_q(j/n_q) \right) \leq \sum_{j=1}^{\lceil \beta n_q \rceil} n_q \cdot \left( 1 + \frac{1}{n} + \sum_{k=1}^{q+1} \frac{(j/n_q + 2q/n)^k}{k!} \right)$$

$$= \sum_{j=1}^{\lceil \beta n_q \rceil} \left( \frac{n_q}{n} + n_q \right) + \sum_{k=1}^{q+1} \sum_{j=1}^{\lceil \beta n_q \rceil} \frac{n_q}{k!} \cdot \left( \frac{j}{n_q} + \frac{2q}{n} \right)^k .$$

The former sum can be bounded using $\beta \leq 1$, the latter one — using Lemma 5:

$$T_{q+1}(\beta) \leq \frac{n_q^2}{n} + (\beta n_q + 1) \cdot n_q + \sum_{k=1}^{q+1} \frac{n_q^2}{(k+1)!} \cdot \left( \beta + \frac{2(q+1)}{n} \right)^{k+1}$$

$$= \frac{n_{q+1}}{n} + n_{q+1} \cdot (\beta + 1/n_q) + n_{q+1} \cdot \sum_{k=2}^{q+2} \frac{(\beta + 2(q+1)/n)^k}{k!}$$

$$\leq \frac{n_{q+1}}{n} + n_{q+1} \cdot \sum_{k=1}^{q+2} \frac{(\beta + 2(q+1)/n)^k}{k!} .$$

In the last inequality, we used $1/n_q \leq 2(q+1)/n$. This completes the proof. □

## 4.2 Bounding the Efficiency of $S_q(\beta)$

In this section, we measure the efficiency of routines $S_q(\beta)$ in reducing the load in a set of buffers. Fix any set $A$ of $n_q$ full buffers. We would like to define $D_q(\beta)$ as the best achievable upper bound on the load in $A$ after executing strategy $S_q(\beta, A)$. However, such definition would be useless as the algorithm may decide to transmit only from $\mathcal{B} \setminus A$, keeping the buffers of $A$ full. Hence, we adapt the formulation that takes such behavior into account.

**Definition 7.** *For a given integer $q \geq 0$ and real $\beta \in (0, 1]$, we define the* remainder $D_q(\beta)$ *of the routine $S_q(\beta)$ as the infimum of all values of $D$ for which the following statement holds: "If at time $\tau$* there is a subset $A$ of $n_q$ full buffers, and the adversary starts the strategy $S_q(\beta, A)$, then for any load $H$ transmitted by an algorithm during this strategy from $\mathcal{B} \setminus A$, either (i) the total load drops below 1 during $S_q(\beta, A)$, or (ii) the load in $A$ at time $(\tau + T_q(\beta))^+$ is at most $H + D''$.

For example, by Lemma 3, $D_0(\beta) \leq n \cdot e^{-\beta} + 1$. Throughout the rest of this section, we denote $\exp(-1/n_q)$ by $\gamma$. First, we present the recursive relation between remainders.

**Lemma 8.** *Fix any integer $q \geq 0$ and a real $\beta \in (0, 1]$. Let $w = \lceil \beta n_q \rceil$. Then,*

$$D_{q+1}(\beta) \leq n_{q+1} \cdot \gamma^w + \sum_{j=1}^{w} \gamma^{w-j} \cdot \left( D_q(j/n_q) - n_q + 1 \right) \ .$$

*Proof.* We choose any time $\tau$, any set $A$ of $n_{q+1} = n_q^2$ full buffers and any situation at the remaining buffers, and analyze the strategy $S_{q+1}(\beta, A)$ starting at time $\tau$. We may assume that during the execution of $S_{q+1}(\beta, A)$ the total load never drops below 1; otherwise the lemma follows trivially. By Definition 4, $S_{q+1}(\beta, A)$ consists of $w$ phases, where the $j$-th phase is an action $(A, n_q)$ affecting set $A_j$ of $n_q$ buffers, followed by the strategy $S_q(j/n_q, A_j)$.

Let $H$ be the load transmitted from buffers $\mathcal{B} \setminus A$ during strategy $S_{q+1}(\beta, A)$. Moreover, for the $j$-th phase, we define:

- $H_j$: the load transmitted during action $(A, n_q)$ from the buffers $\mathcal{B} \setminus A$;
- $H'_j$: the load transmitted during $S_q(j/n_q, A_j)$ from the buffers $\mathcal{B} \setminus A$;
- $H''_j$: the load transmitted during $S_q(j/n_q, A_j)$ from the buffers $A \setminus A_j$.

Clearly, $H = \sum_{j=1}^{w}(H_j + H'_j)$.

Let $t_0 = \tau$ and let $t_j$ be the ending time of phase $j$, i.e., phase $j$ lasts between $t_{j-1}$ and $t_j$. For $j \in \{0, \ldots, w\}$, let $C_j$ denote the load in $A$ at time $t_j^+$, i.e., $C_j$ is the load in $A$ at the end of phase $j$. Clearly, $C_0 = n_{q+1}$ as all the buffers of $A$ are full at the beginning. Our goal is to upper-bound the load in buffers from $A$ when the strategy $S_{q+1}(\beta, A)$ terminates, i.e., the value of $C_w$. To this end, we focus on a single phase $j$ and relate $C_j$ to $C_{j-1}$.

The action $(A, n_q)$ takes place in steps $(t_{j-1}, t_{j-1} + n_q]$. Since the algorithm is work-conserving and the total load never drops below 1, the algorithm transmits the load $n_q - H_j$ from the buffers of $A$ during these steps. The load of $A$ at time $(t_{j-1} + n_q)^-$ is then $C_{j-1} - n_q + H_j$. At time $t_{j-1} + n_q$, the adversary chooses set $A_j$ of $n_q$ most populated buffers of $A$ and injects a packet to each buffer of $A_j$. Next, we consider the load in buffers from $A \setminus A_j$ and $A_j$ separately.

- As $A_j$ were the most loaded $n_q$ buffers from $A$ at time $(t_{j-1} + n_q)^-$, the load in $A \setminus A_j$ at time $(t_{j-1} + n_q)^+$ is at most $(C_{j-1} - n_q + H_j) \cdot (1 - n_q/n_{q+1})$. Recall that $H''_j$ is the load transmitted from the buffers of $A \setminus A_j$ during the strategy $S_q(j/n_q, A_j)$. Thus, the load in $A \setminus A_j$ at time $t_j^+$ is at most $(C_{j-1} - n_q + H_j) \cdot (1 - n_q/n_{q+1}) - H''_j$.

- All buffers from $A_j$ are full at time $(t_{j-1} + n_q)^+$. In steps $(t_{j-1} + n_q, t_j]$, the adversary executes the strategy $S_q(j/n_q, A_j)$. By the inductive assumption, the load in $A_j$ at time $t_j^+$ is at most $D_q(j/n_q) + (H'_j + H''_j)$.

In total, the load in $A$ at time $t_j^+$ is

$$\begin{aligned} C_j &\leq \left(C_{j-1} - n_q + H_j\right) \cdot (1 - 1/n_q) - H''_j + D_q(j/n_q) + H'_j + H''_j \\ &\leq C_{j-1} \cdot \gamma - n_q + 1 + D_q(j/n_q) + (H_j + H'_j) \ . \end{aligned} \quad (1)$$

We obtain a closed-form formula for $C_w$ by solving the recurrence (1):

$$C_w \leq C_0 \cdot \gamma^w + \sum_{j=1}^{w} \gamma^{w-j} \cdot \left(D_q(j/n_q) - n_q + 1 + H_j + H'_j\right) .$$

Finally, substituting $C_0 = n_{q+1}$ and using $\gamma < 1$,

$$C_w \leq H + n_{q+1} \cdot \gamma^w + \sum_{j=1}^{w} \gamma^{w-j} \cdot (D_q(j/n_q) - n_q + 1) .$$

As $C_w$ is the load in $A$ when $S_{q+1}(\beta, A)$ terminates, the lemma follows. □

**Lemma 9.** *For any integer $q \geq 0$ and real $\beta \in (0, 1]$, it holds that $D_q(\beta) \leq n_q \cdot (e^{-\beta} \cdot F_q(\beta) + Q_q)$, where*

$$F_q(\beta) = \sum_{h=0}^{q} \sum_{k=0}^{h} \frac{(\beta + 2q/n)^k}{k!} \quad \text{and} \quad Q_q = -q + \frac{q+1}{n} .$$

*Proof.* The proof is by induction on $q$. As $n_0 = n$, $Q_0 = 1/n$ and $F_0(\beta) = 1$, the induction basis ($q = 0$) states that $D_0(\beta) \leq (n \cdot e^{-\beta} + 1)$, which holds by Lemma 3. In the following, we fix $q$, assume that $D_q(\beta) \leq n_q \cdot (e^{-\beta} \cdot F_q(\beta) + Q_q)$ for any $\beta \in (0, 1]$, and we show an appropriate bound on $D_{q+1}(\beta)$ (also for any $\beta \in (0, 1]$). Let $w = \lceil \beta n_q \rceil$. By Lemma 8, and further applying the inductive assumption

$$\begin{aligned} D_{q+1}(\beta) &\leq n_{q+1} \cdot \gamma^w + \sum_{j=1}^{w} \gamma^{w-j} \cdot \left(D_q(j/n_q) - n_q + 1\right) \\ &\leq n_{q+1} \cdot \gamma^w + \sum_{j=1}^{w} \gamma^{w-j} \cdot n_q \gamma^j F_q(j/n_q) + \sum_{j=1}^{w} \gamma^{w-j} \cdot \left(n_q Q_q - n_q + 1\right) . \end{aligned} \quad (2)$$

We focus on upper-bounding the second sum. As $n \leq n_q$, by the definition of $Q_{q+1}$, it holds that $n_q \cdot Q_q - n_q + 1 \leq n_q \cdot Q_{q+1}$. As the term $n_q \cdot Q_{q+1}$ is negative, we need to lower-bound the expression $\sum_{j=1}^{w} \gamma^{w-j} = (1 - \gamma^w)/(1 - \gamma)$. To this end, we note that $\gamma = \exp(-1/n_q) \geq 1 - 1/n_q$, and equivalently $1/(1 - \gamma) \geq n_q$. Hence, the second sum is at most $n_q \cdot (1 - \gamma^w) \cdot n_q \cdot Q_{q+1} \leq n_{q+1} \cdot (1 - \gamma^w) \cdot Q_{q+1}$. Plugging the achieved bound to (2), we obtain

$$\begin{aligned} D_{q+1}(\beta) &\leq n_{q+1} \cdot \gamma^w + n_q \cdot \sum_{j=1}^{w} \gamma^w \cdot F_q(j/n_q) + n_{q+1} \cdot (1 - \gamma^w) \cdot Q_{q+1} \\ &\leq n_{q+1} \cdot \gamma^w \cdot \left(1 - Q_{q+1} + \frac{1}{n_q} \sum_{j=1}^{w} F_q(j/n_q)\right) + n_{q+1} \cdot Q_{q+1} . \end{aligned} \quad (3)$$

We denote the expression in brackets by $E$. To bound it, we use $w = \lceil \beta n_q \rceil$ and substitute the values of $F_q(j/n_q)$. Next, we observe that $-Q_{q+1} \leq q + 1$ and use Lemma 5, obtaining

$$\begin{aligned}
E &= 1 - Q_{q+1} + \frac{1}{n_q} \sum_{j=1}^{\lceil \beta n_q \rceil} \sum_{h=0}^{q} \sum_{k=0}^{h} \frac{(j/n_q + 2q/n)^k}{k!} \\
&\leq q + 2 + \sum_{h=0}^{q} \sum_{k=0}^{h} \frac{(\beta + 2(q+1)/n)^{k+1}}{(k+1)!} = \sum_{h=0}^{q+1} \sum_{k=0}^{h} \frac{(\beta + 2(q+1)/n)^k}{k!} \\
&= F_{q+1}(\beta) \ .
\end{aligned} \qquad (4)$$

Combining (4) with (3) yields $D_{q+1}(\beta) \leq n_{q+1} \cdot \gamma^w \cdot F_{q+1}(\beta) + n_{q+1} \cdot Q_{q+1}$. Applying $\gamma^w = \exp(-\lceil \beta n_q \rceil / n_q) \leq \exp(-\beta)$ yields $D_{q+1}(\beta) \leq n_{q+1} \cdot (e^{-\beta} \cdot F_{q+1}(\beta) + Q_{q+1})$, which completes the inductive proof. $\square$

### 4.3 The Competitive Ratio

The remaining part of our reasoning is to verify the performance of the routine $S_q(1)$ executed on a set of $n_q$ initially full buffers, where $n_0 = n$ will be chosen to be sufficiently larger than $q$. As our bounds on the efficiency of $S_q(1)$ hold provided the total load of an algorithm does not drop below 1, we slightly adapt our strategy.

**Definition 10.** *For any integer $q \geq 0$, the adversarial strategy $S'_q(A)$ on a set $A$ of $n_q$ buffers is to fill these buffers at the beginning and then run strategy $S_q(1, A)$ on them. However, if a load of an algorithm drops below 1, the adversary finishes only the current action, and then ends the strategy immediately.*

**Lemma 11.** *Fix any integer $q \geq 1$ and let $n = q^4$. Fix any online algorithm ALG. Assume that $m \geq n_q$ and choose any subset $A$ of $n_q$ buffers. If the adversary executes strategy $S'_q(A)$, then the OPT-to-ALG throughput ratio is at least $e/(e-1) - O(1/q)$.*

*Proof.* In our proof we use the two following bounds:

$$(1 + 2q/n)^{q+1} < e^{(2q/n)\cdot(q+1)} \leq e^{4/q^2} = 1 + O(q^{-2}) \qquad (5)$$

and

$$\begin{aligned}
\sum_{h=0}^{q} \sum_{k=0}^{h} \frac{1}{k!} &= \sum_{k=0}^{q} (q + 1 - k) \cdot \frac{1}{k!} = (q + 1) \cdot \sum_{k=0}^{q} \frac{1}{k!} - \sum_{k=0}^{q-1} \frac{1}{k!} \\
&= \frac{1}{q!} + q \cdot \sum_{k=0}^{q} \frac{1}{k!} < \frac{1}{q!} + q \cdot e \ .
\end{aligned} \qquad (6)$$

Let $\ell$ be the last time when packets were injected. If the strategy does not end prematurely (because of the total load dropping below 1), then $\ell = T_q(1)$, and otherwise $\ell \leq T_q(1)$. In either case, we bound the value of $T_q(1)$ using Lemma 6 and (5), obtaining

$$\begin{aligned}
\ell &\leq \left( \frac{1}{n} + \sum_{k=1}^{q+1} \frac{(1 + 2q/n)^k}{k!} \right) \cdot n_q < \left( \frac{1}{n} + \left(1 + O(q^{-2})\right) \cdot \sum_{k=1}^{q+1} \frac{1}{k!} \right) \cdot n_q \\
&< \left( e - 1 + O(q^{-2}) \right) \cdot n_q \ .
\end{aligned} \qquad (7)$$

Let $c$ be the total load at time $\ell^+$. Assume first that a sequence has ended prematurely, i.e., the total load has dropped below 1 during some action of length $a$. By the construction of the routine $S_q(1)$, all its actions have length at most $n_{q-1}$. Therefore, $a \leq n_{q-1}$ packets are injected at time $\ell$, and hence $c < 1 + n_{q-1} = O(1/n_{q-1}) \cdot n_q = O(1/q) \cdot n_q$.

Now, we consider the case when the strategy $S_q(1, A)$ was executed completely. As the buffers $\mathcal{B} \setminus A$ are always empty, the algorithm can only transmit from buffers inside $A$, i.e, the term $H$ in Definition 7 is zero. Hence, in this case $c \leq D_q(1)$. To compute the value of $D_q(1)$, we use the bound of Lemma 9, and further apply (5) and (6):

$$\begin{aligned}
c &\leq \left( e^{-1} \cdot \sum_{h=0}^{q} \sum_{k=0}^{h} \frac{(1 + 2q/n)^k}{k!} - q + \frac{q+1}{n} \right) \cdot n_q \\
&\leq \left( e^{-1} \cdot (1 + O(q^{-2})) \cdot \sum_{h=0}^{q} \sum_{k=0}^{h} \frac{1}{k!} - q + \frac{q+1}{n} \right) \cdot n_q \quad (8) \\
&< \left( (1 + O(q^{-2})) \cdot \left( q + \frac{1}{e \cdot q!} \right) - q + \frac{q+1}{n} \right) \cdot n_q \\
&= O(1/q) \cdot n_q \ .
\end{aligned}$$

The throughput of ALG is at most $\ell + c$. On the other hand, the strategy $S'_q(A)$ is canonical, and hence by Lemma 1, the throughput of OPT is $\ell + n_q$. Therefore, by (7) and (8), the OPT-to-ALG throughput ratio is at least $(\ell + n_q)/(\ell + c) = e/(e-1) - O(1/q)$. □

**Theorem 12.** *The competitive ratio of any fractional algorithm for the unweighted FIFO variant of the buffer management problem is at least $e/(e-1)$. The bound holds for any value of $B$ and $m \to \infty$.*

*Proof.* We show that for any $m \geq 16$ and any $B$, the lower bound on the competitive ratio is $e/(e-1) - O(1/\log \log m)$. To this end, we pick $q = \lfloor \frac{1}{2} \log \log m \rfloor$. Then choosing $n = q^4$, we obtain

$$n_q = n^{2^q} = q^{4 \cdot 2^q} = 2^{2^q \cdot 4 \log q} \leq 2^{4^q} \leq 2^{4^{(\log \log m)/2}} = m \ ,$$

and hence applying the routine $S'_q$ is feasible. By Lemma 11, the ratio is at least $e/(e-1) - O(1/q) = e/(e-1) - O(1/\log \log m)$, as claimed. This bound becomes arbitrarily close to $e/(e-1)$, when $m$ grows. □

## 5 Fractional vs. Randomized Algorithms

The following straightforward reduction shows that the fractional model is easier for algorithms than the randomized one. The lemma below together with Theorem 12 yield the lower bound of $e/(e-1)$ on the performance of any randomized algorithm.

**Lemma 13.** *For any randomized algorithm RAND for the integral model with competitive ratio $R$, there exists a deterministic algorithm FRAC for the fractional model whose ratio is at most $R$ (for the same buffer size and number of buffers).*

*Proof.* We fix any buffer size $B$ and number of buffers $m$.

For an algorithm $A$, we denote the number of packets in its $i$-th buffer by $Q_A(i)$. The proof follows by simulation, i.e., we show that on the basis of RAND, it is possible to construct an algorithm FRAC, such that — at any time of any input sequence — the following invariant holds for any buffer $i$:

$$Q_{\text{FRAC}}(i) \geq \mathbf{E}[Q_{\text{RAND}}(i)] \ ,$$

and for any step where RAND transmits $p$ packets in expectation, FRAC transmits load $p$ as well. This implies that — on any input sequence — the throughput of FRAC is at least that of RAND, which yields the lemma.

Clearly, the invariant holds at the beginning, when both RAND and FRAC have no packets. Then, the proof follows inductively, by considering two possible types of events.

- A packet is injected to buffer $i$. If $Q_{\text{FRAC}}(i) \geq B - 1$ before the injection, then afterwards $Q_{\text{FRAC}}(i) = B$. Otherwise, $Q_{\text{FRAC}}(i)$ increases by 1 and $\mathbf{E}[Q_{\text{RAND}}(i)]$ increases at most by 1. In both cases, the invariant is preserved.

- Packets are transmitted during a step. Let $Q'_{\text{RAND}}(i)$ be the number of packets in buffer $i$ after the transmission of RAND. Let $q_i = \mathbf{E}[Q_{\text{RAND}}(i) - Q'_{\text{RAND}}(i)]$. Clearly, $q_i \geq 0$ for all $i$ and $\sum_{i=1}^{m} q_i \leq 1$. To simulate this behavior of RAND and preserve the invariant, FRAC transmits load $q_i$ from buffer $i$. □

**Corollary 14.** *The competitive ratio of any randomized algorithm for the unweighted FIFO variant of the buffer management problem on is at least $e/(e-1)$. The bound holds for any value of B and $m \to \infty$.*

By means of Lemma 13, lower bounds for the fractional model apply to randomized algorithms. Below, we show that for the upper bounds it is not necessarily the case, i.e., it is not possible to perform online randomized rounding of a fractional solution (for $m \geq 3$) that preserves the throughput. This may be an indication that the randomized and the fractional model are not equivalent in terms of the competitive ratio.

**Theorem 15.** *For $m \geq 3$ online randomized rounding of a fractional solution that preserves a throughput of the algorithm is not feasible.*

*Proof.* We show how to create an input sequence and a fractional solution FRAC, such that any randomized algorithm RAND that tries to simulate FRAC has smaller throughput. We assume that $m = 3$. (If $m > 3$ nothing is ever injected to the additional buffers, and hence both fractional and randomized algorithm have them empty.)

As in the proof of Lemma 13, for an algorithm $A$, we denote the number of packets in its $i$-th buffer by $Q_A(i)$. We say that RAND *exactly simulates* FRAC at time $t$ if for any buffer $i$ it holds that $\mathbf{E}[Q_{\text{RAND}}(i)] = Q_{\text{FRAC}}(i)$. We first argue that this condition is necessary for preserving the throughput.

Precisely speaking, assume that at time $t^+$, RAND exactly simulates FRAC and their throughput has been equal so far. Assume that within time interval $(t, t + k)$, no packet is injected and FRAC transmits load $k$. If RAND does not exactly simulate FRAC at time $(t+k)^-$, then its total throughput can be made lower than that of FRAC. To show this claim, we consider two cases.

1. If the expected load transmitted by RAND in steps $(t, t + k)$ is smaller than $k$, then the adversary may inject $B$ packets to each of the buffers at time $t + k$ and end the input sequence afterwards. Hence, the throughput of RAND in step $(t, t + k)$ is smaller than that of FRAC and the throughput in the remaining steps is equal to $m \cdot B$ for both algorithms.

2. If the expected load transmitted by RAND in time steps $(t, t + k)$ is equal to $k$, then $\sum_{i=1}^{m} \mathbf{E}[Q_{\text{RAND}}(i)] = \sum_{i=1}^{m} Q_{\text{FRAC}}(i)$. As the exact simulation condition is violated, there exists a buffer $i$ for which $\mathbf{E}[Q_{\text{RAND}}(i)] < Q_{\text{FRAC}}(i)$. In this case, at time $t + k$, the adversary injects $B$ packets to all buffers but the $i$-th one and ends the input sequence. The throughput of FRAC in the remaining steps is then $(m-1) \cdot B + Q_{\text{FRAC}}(i)$ while the expected throughput of RAND is strictly smaller, i.e., $(m-1) \cdot B + \mathbf{E}[Q_{\text{RAND}}(i)]$.

It remains to construct an input sequence and a behavior of Frac that cannot be simulated exactly by Rand. We identify the state of an algorithm with the state of its buffers. In these terms, the state of Frac at a given time $t$ is a triplet denoted $L(t)$. At time 0, the adversary fills all three buffers with packets, which means that $L(0^+) = (B, B, B)$. Clearly, the state of Rand is the same.

In step $(0, 1)$, Frac transmits a half of the packet from the first and the second buffer, i.e., $L(1^-) = (B-1/2, B-1/2, B)$. To assure an exact simulation, Rand has to transmit with probability $1/2$ from the first buffer and with the probability $1/2$ from the second one, ending with state $(B-1, B, B)$ with probability $1/2$ and in state $(B, B-1, B)$ with probability $1/2$. We denote this distribution over possible states by $\{(B-1, B, B)^{1/2}, (B, B-1, B)^{1/2}\}$ for short. At time 1, the adversary injects a packet to the second buffer, i.e., $L(1^+) = (B-1/2, B, B)$. The probability distribution of Rand simply changes to $\{(B-1, B, B)^{1/2}, (B, B, B)^{1/2}\}$.

In step $(1, 2)$, Frac transmits a half of the packet from the second and the third buffer, i.e., $L(2^-) = (B-1/2, B-1/2, B-1/2)$. How can Rand assure an exact simulation? Let $p_i$ be the probability of transmitting from buffer $i$ conditioned on Rand being in state $(B-1, B, B)$. Similarly, let $q_i$ be the probability of transmitting from buffer $i$, conditioned on Rand being in state $(B, B, B)$. Clearly, $p_1 + p_2 + p_3 = 1$ and $q_1 + q_2 + q_3 = 1$. Furthermore, $p_1 = q_1 = 0$, as the expected load in the first buffer is $B-1/2$ already at time $1^+$. As the expected load in the two remaining buffers has to be also $B-1/2$, it follows that $1/2 \cdot (1-p_i) + 1/2 \cdot (1-q_i) = 1/2$ for $i \in \{2, 3\}$. Hence, $p_2 + q_2 = 1$ and $p_3 + q_3 = 1$. For any fixed $p_2$, at step $2^-$, the probability distribution of Rand is then $\{(B-1, B-1, B)^{p_2/2}, (B-1, B, B-1)^{(1-p_2)/2}, (B, B-1, B)^{(1-p_2)/2}, (B, B, B-1)^{p_2/2}\}$. The adversary does not inject a packet at step 2, and thus this is also the probability distribution of Rand at step $2^+$.

Now we consider two cases. If $p_2 \geq 1/2$, then Frac empties its first and its second buffer in steps $(2, 2B+1)$, so that $L((2B+1)^-) = (0, 0, B-1/2)$. For the exact simulation Rand has to transmit $2B-1$ packets in these steps and is not allowed to transmit anything from the third buffer, independently of its actual state at time $2^+$. However, with probability $p_2/2 \geq 1/4$, Rand is at state $(B-1, B-1, B)$ at time $2^+$, which renders the exact simulation impossible. Similarly, for the case $p_2 < 1/2$, Frac empties the first and the third buffer in steps $(2, 2B+1)$, and Rand cannot simulate this exactly as at time $2^+$ it is in state $(B-1, B, B-1)$ with probability $(1-p_2)/2 > 1/4$. □

## 6 Flaw in the Analysis of the Random Permutation Algorithm

The lower bound presented in this paper contradicts the claimed performance of the Random Permutation algorithm. In this section, we demonstrate an error in its original analysis [Sch05]. Random Permutation is defined for $B = 1$ and any number $m$ of buffers. It chooses a permutation of the buffers uniformly at random. In any time step, it transmits a packet from the populated buffer which is most to the front in the permutation.

The first packet injected to a particular buffer is called *initializing* and all the subsequent packets injected to this buffer are *non-initializing*. All non-initializing packets are numbered in the order of their arrival and the probability of accepting the $i$-th non-initializing packet is denoted $p_i$. Let

$$v_i = \frac{1}{m!} \cdot \sum_{j=1}^{i} \binom{i}{j} \cdot (m-j)! \cdot (-1)^{j-1} \ . \tag{9}$$

Lemma 1 of [Sch05] states that $p_i \geq v_i$ for $1 \leq i \leq m$, provided that Opt can accommodate all non-initializing packets.

The proof of this lemma considers first an adversarial strategy, where all $m$ initializing packets are injected at time 0 and then for $1 \leq i \leq m$, the $i$-th non-initializing packet is injected at time $i$ to buffer $i$. We call such a strategy *systematic*. For systematic strategies, a recursive formula for $p_i$ shows that indeed $p_i = v_i$ [Sch05].

However, in the remaining part of the proof of [Sch05], it is (informally) argued that for other adversarial strategies the values of $p_i$ can be only higher, i.e., that $p_i \geq v_i$ for $1 \leq i \leq m$. This statement can be falsified immediately by the observation on delayed injections given in Section 1.4: it is possible to make a single value of $p_i$ smaller than $1/2$, whereas $v_m$ tends to $1 - 1/e \approx 0.632$ when $m$ grows.

The subsequent parts of the proof of the RANDOM PERMUTATION competitiveness use only the weaker relation $\sum_{i=1}^{m} p_i \geq \sum_{i=1}^{m} v_i$. However, the delayed injections can be employed to show that even such a relation is false.

**Lemma 16.** *There exists an integer $m$ and the sequence of adversarial injections of $m$ initializing and $m$ non-initializing packets leading to $\sum_{i=1}^{m} p_i < \sum_{i=1}^{m} v_i$.*

*Proof.* All initializing packets are injected at time 0. For $1 \leq i \leq m - 3$, the $i$-th non-initializing packet is injected to buffer $i$ at time $i$. At time $m - 2$, no packets are injected. At time $m - 1$, the adversary chooses two buffers, denoted $b_1$ and $b_2$, with the maximum expected number of packets, and injects a packet to both these buffers. At time $m$, the adversary injects a packet either to $b_1$ or to $b_2$, choosing the buffer maximizing the expected number of packets.

Below, we show that for $m = 16$, it holds that $\sum_{i=1}^{m} p_i < \sum_{i=1}^{m} v_i$. Note that up to step $m - 3$, the injection pattern is systematic, and thus $\sum_{i=1}^{m-3} p_i = \sum_{i=1}^{m-3} v_i$. It is therefore sufficient to show that $\sum_{i=m-2}^{m} p_i < \sum_{i=m-2}^{m} v_i$.

In each step $(i - 1, i]$ for $i \leq m - 3$, RANDOM PERMUTATION transmits a packet and accepts an expected number of $p_i$ packets. Hence, the total expected load at time $(m - 3)^+$ is $m + \sum_{i=1}^{m-3}(-1 + p_i) = 3 + \sum_{i=1}^{m-3} v_i$. The load at time $(m - 1)^-$ is then $1 + \sum_{i=1}^{m-3} v_i$, and therefore the total expected number of packets in buffers $b_1$ and $b_2$ is at least $\frac{2}{m} \cdot (1 + \sum_{i=1}^{m-3} v_i)$. Recall that $p_{m-2} + p_{m-1}$ is the expected number of packets accepted to these two buffers at time $m - 1$. Thus,

$$p_{m-2} + p_{m-1} \leq 2 - \frac{2}{m} \cdot \left(1 + \sum_{i=1}^{m-3} v_i\right) . \tag{10}$$

As at time $(m - 1)^+$ the buffers $b_1$ and $b_2$ are full, at time $m^-$ the expected amount of packets in the more loaded buffer is at least $1/2$, and hence $p_m \leq 1/2$. Combining this bound with (10) yields

$$\sum_{i=m-2}^{m} p_i \leq \frac{5}{2} - \frac{2}{m} - \frac{2}{m} \cdot \sum_{i=1}^{m-3} v_i . \tag{11}$$

Taking $m = 16$, calculating the actual values of $v_i$, and plugging them into the right hand side of (11) yields

$$\sum_{i=m-2}^{m} p_i \leq \frac{609\,769\,643\,705\,700}{m \cdot m!} < \frac{611\,230\,500\,168\,224}{m \cdot m!} = \sum_{i=m-2}^{m} v_i ,$$

which finishes the proof. $\square$